# Afterpulsing and Instability in Superconducting Nanowire Avalanche Photodetectors

F. Marsili[1], F. Najafi[1], E. Dauler[2], R. J. Molnar[2], K. K. Berggren[1*]

[1] Department of Electrical Engineering and Computer Science, Massachusetts Institute of Technology, 77 Massachusetts Avenue, Cambridge, Massachusetts 02139, USA

[2] Lincoln Laboratory, Massachusetts Institute of Technology, 244 Wood St., Lexington, Massachusetts 02420, USA

We investigated the reset time of superconducting nanowire avalanche photodetectors (SNAPs) based on 30 nm wide nanowires. We studied the dependence of the reset time of SNAPs on the device inductance and discovered that SNAPs can provide a speed-up relative to SNSPDs with the same area, but with some limitations: (1) reducing the series inductance of SNAPs (necessary for the avalanche formation) could result in the detectors operating in an unstable regime; (2) a trade-off exists between maximizing the bias current margin and minimizing the reset time of SNAPs; and (3) reducing the reset time of SNAPs below ~ 1 ns resulted in afterpulsing.

Superconducting nanowire avalanche photodetectors (SNAPs, also referred to as cascade-switching superconducting single-photon detectors [1]) are based on a parallel-nanowire architecture that performs single-photon counting with signal-to-noise ratio ($SNR$) up to a factor of ~ 4 higher [2] than traditional superconducting nanowire single-photon detectors (SNSPDs) [3, 4]. Although we recently worked to improve our understanding of the operation mechanism of SNAPs [2, 5], the claim that these devices can operate at higher speed than SNSPDs [1] has not yet been confirmed experimentally. We studied the reset time of SNAPs and found that although SNAPs can provide a speed-up relative to SNSPDs with the same area, the device speed is limited by the thermal relaxation of the nanowires.

We investigated the possibility of reducing the reset time of SNAPs below 1 ns by varying the number of sections in parallel ($N$) and by decreasing the device series inductance ($L_S$) necessary for the avalanche formation [1]. Indeed, as the equivalent kinetic inductance of $N$ nanowires in parallel is $N^2$ times lower than their inductance in series, a $N$-parallel-section SNAP ($N$-SNAP) would be $N^2$ times faster than a SNSPD of the same active area, if one were to neglect certain non-idealities of the device. SNAPs operate by having the current from the section which switches to the normal state after absorbing a photon (initiating section) drive the still-superconducting sections

[*] corresponding author: berggren@mit.edu

(secondary sections) normal, resulting in a current redistribution to the read-out (modeled as a resistor $R_{load}$) [2, 5]. Ideally, no current from the initiating section should be diverted to $R_{load}$ before the secondary sections have switched to the normal state (we call this ideal operation mechanism *perfect redistribution* [2]). However, in practice the current leaking to $R_{load}$ (we call this current *leakage current*, $I_{lk}$) can be substantial. The series inductor is used to minimize the leakage current and it has a dominant effect on the device speed [1].

We studied the effect of reducing $L_S$ on the device operation by introducing a unitless parameter $r$, which we defined as the ratio between $I_{lk}$ and the current redistributing to all the secondary sections after the initiating section switches to the normal state. For perfect redistribution, $r = 0$. Considering times much shorter than the reset time (so that the inductive impedance dominates the read-out resistance, which is typically $R_{load} = 50\ \Omega$): $r = L_0 / [L_S \cdot (N - 1)]$ [6], where $L_0$ is the kinetic inductance of one section. Therefore, when decreasing $L_S$ (and thus the detector reset time), the leakage current increases and a higher bias current is necessary to ensure an avalanche.

We characterized ~ 100 $N$-SNAPs with $N = 2$, 3 and 4 and with $r$ ranging from 0.1 to 2 [7] by measuring the photoresponse count rate and the photoresponse inter-arrival time histograms (see supplementary material [8] and Ref. [2] for details of the fabrication process, detector geometry, and experimental setup). We found that, depending on $r$ and on the bias current, devices could exhibit (1) correct operation as single-photon detectors (avalanche regime [2]); (2) operation in arm-trigger regime[2]; (3) unstable operation; or (4) after-pulsing.

An unstable operating regime of SNAPs was observed in devices with low $L_S$ ($r > 0.1$) biased at low current, before the onset of the arm-trigger regime. In the unstable regime, after a hotspot nucleation (HSN) event (caused either by the absorption of a photon or by a dark count) occurred in one of the sections of the SNAP, the device emitted multiple current pulses. These trains of pulses resulted in a spurious peak in the normalized photoresponse count rate (*PCR*) vs normalized bias current ($I_B / I_{SW}$, where $I_{sw}$ is the device switching current, defined as the highest $I_B$ the device was able to sustain before switching to the normal state) curves, as shown in Figure 1a for a 3-SNAP. As the *PCR* in this operating regime was only weakly dependent on the photon flux incident on the device (increasing the photon flux by a factor of ~ 15 changed the count rate by a factor of ~ 4 [9]), normalization by the applied photon flux (see supplementary material [8] for details) meant that the amplitude of the spurious peak decreased with increasing optical power. The normalized *PCR* vs $I_B$ curves indicated that the devices transitioned through three operating regimes as the bias current was increased: (1) the unstable regime ($0.62 \lesssim I_B / I_{SW} \lesssim 0.7$); (2) the arm-trigger regime ($0.7 \lesssim I_B / I_{SW} \lesssim 0.82$); and (3) the avalanche regime ($I_B / I_{SW} \gtrsim 0.82$). The devices worked as single-photon detectors only when operating in the avalanche regime, in which the normalized *PCR* could then be identified with the detection efficiency. Indeed, in the arm-trigger regime two or more HSN events were necessary to produce a count [2] and in the unstable regime one single HSN event produced multiple counts. The inset of Figure 1a shows the oscilloscope traces of a detector response in the unstable (blue curve) and the avalanche regimes (red curve). In the unstable regime, pulses with two distinct average amplitudes were recorded (which we called "small" and "large" pulses [10]). Unlike the arm-trigger and avalanche regimes (discussed in Ref. [2]), the unstable regime was not previously observed, so we will discuss it in detail here.

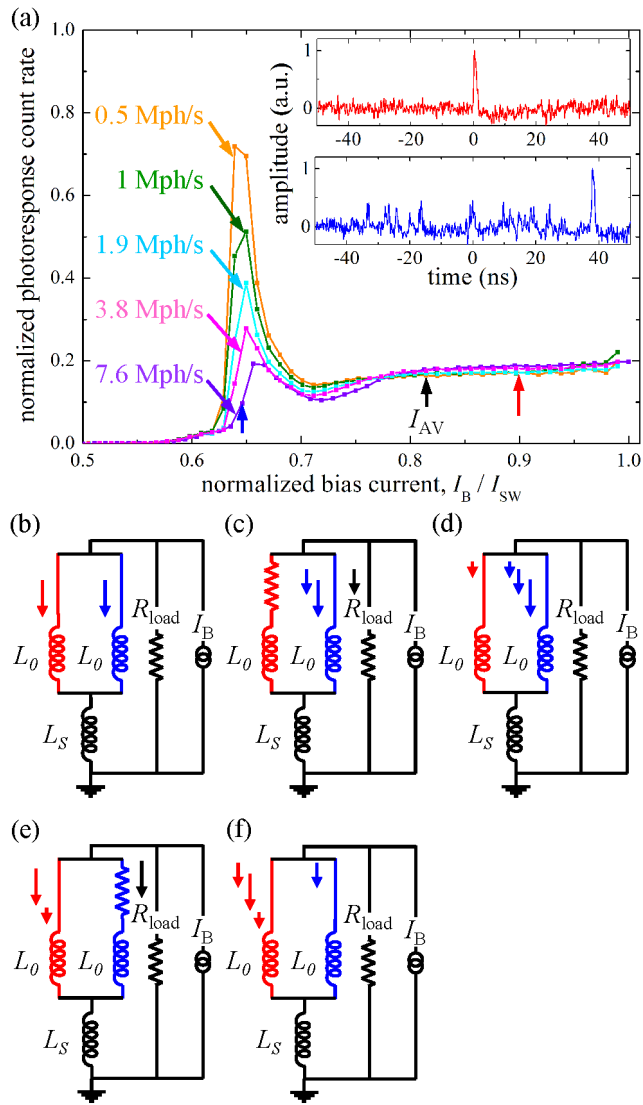

**Figure 1. a.** Normalized *PCR* vs normalized $I_B$ of a 30-nm-wide 3-SNAP (1.47 μm × 830 nm active area) at different photon fluxes (expressed in photons per second: ph/s). The *PCR* is normalized to the photon flux (see supplementary material in Ref. 8). The kinetic inductance of one section was $L_0$ = 13 nH (estimated from the fall time of the detector response pulse) and the value of the series inductor was $L_S$ = 1.8 $L_0$ ($r$ = 0.28).

The detector avalanche current ($I_{AV}$), marked by a black arrow, was determined experimentally as reported in Ref. 2. **Inset.** Oscilloscope traces of the photoresponse of the 3-SNAP of Figure 1a measured in the unstable ($I_B = 0.65\ I_{SW}$, in blue) and avalanche regimes ($I_B = 0.9\ I_{SW}$, in red). The two bias currents are marked by the blue and red arrow respectively in Figure 1a. **b-f.** Electrical equivalent circuit of the different states of a 2-SNAP operating in the unstable regime.

Figure 1b through f illustrate our explanation for the unstable regime. Figure 1b shows the electrical equivalent of an unstable 2-SNAP in the steady state. After a HSN event occurs in section 1 (in red, see Figure 1c) no avalanche is formed because a large part of the redistributed current leaks into $R_{load}$. Once section 1 switches back to the superconducting state (Figure 1d), the current in each of the SNAP sections increases at the same rate, but from different initial conditions. Indeed, while section 1 is depleted of current, section 2 (in blue) is biased close to $I_{SW}$, as its current was not depleted by the initial HSN event. When the current in section 2 exceeds its $I_{SW}$ and section 2 switches to the normal state (Figure 1e), the resulting current redistribution (Figure 1f) brings the system back to the non-stationary state (section 1 normal, section 2 superconducting) illustrated in Figure 1c. From this point on, the different branches of the circuit continue swapping their current periodically, which causes the small pulses observed experimentally. Large pulses are then generated when an avalanche forms in the device as a result of the occurrence of several subsequent HSN events during the instability cycle. The avalanche stops the instability cycle and restores the device to the stationary condition illustrated in Figure 1b. To support our model of the device operation, we simulated the current dynamics of a 2-SNAP and a 3-SNAP with $r = 1$ by using the electrothermal model described in Ref. [5] and reproduced the unstable regime (see supplementary material [8]).

We reduced the photoresponse fall time of SNAPs (the time constant of the exponential decay of the detector response pulse) by decreasing the device series inductance ($L_S$) as shown in Figure 2a for 4-SNAPs (for $r = 0.125$ to 1). However, the resulting decrease in fall time came at the price of an increased avalanche current ($I_{AV}$) as shown in Figure 2b, where the avalanche currents of the same detectors shown in Figure 2a are marked by colored arrows on the normalized *PCR* vs bias current curves. The values of $I_{AV}$ were determined experimentally as reported in Ref. [2]. A high $I_{AV}$ is undesirable because the bias range in which the devices operate as low-jitter single-photon detectors decreases with increasing $I_{AV}$ (as reported in Ref. [2, 11]). Therefore, we concluded that a trade-off exists between minimizing the reset time and maximizing the bias margin of these devices.

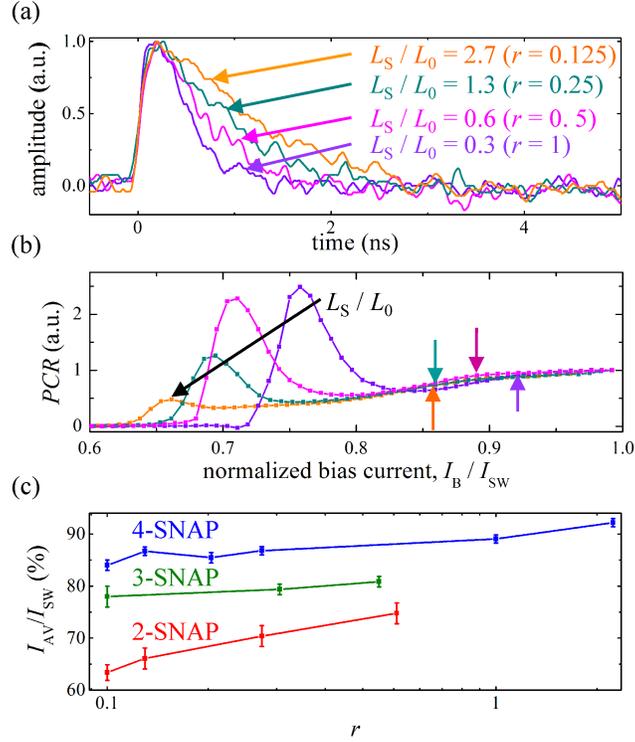

**Figure 2. a.** Single-shot oscilloscope trace of the photoresponse pulses of 30-nm-wide 4-SNAPs (1.47 μm × 1.43 nm active area) with different values of $L_S$. The kinetic inductance of one section was $L_0 = 13$ nH (estimated from the fall time of the detector response pulse). The devices were biased in avalanche regime, at $I_B = 0.98\ I_{SW}$. The waveforms were normalized by their maximum. **b.** PCR (normalized to the PCR at the switching current) vs $I_B / I_{SW}$ of the same 4-SNAPs shown in Figure 2a. The photon flux on the device active area was $2.0 \cdot 10^7$ photons per second. **c.** $I_{AV}$ vs $r$ for 2-, 3- and 4-SNAPs.

Figure 2c shows the dependence of the experimental values of the avalanche current on $N$ and $r$. When characterizing 2-, 3- and 4-SNAPs with decreasing values of $L_S$, we observed an increase in $I_{AV}$ with increasing $r$. The increase in $I_{AV}$ was more pronounced for SNAPs with low $N$, because for a certain value of $I_{lk}$ the current redistributed to each secondary section is $I_{lk} / (N\text{-}1)$ lower than in case of perfect redistribution. The leakage current causes a larger increase in the avalanche current for lower $N$ because the ratio between $I_{lk} / (N\text{-}1)$ and the bias current of each secondary section is larger. This behavior was predicted by an approximate model of the device operation which assumes that all the current of the initiating section is redistributed to the secondary sections and to $R_{load}$ (see supplementary material [8]).

We performed time-resolved measurements to verify whether the decrease in the photoresponse fall time shown in Figure 2a corresponded to an effective decrease in the detector reset time (as the reset time and the photoresponse fall time scale differently with the device inductance in similar detectors [12, 13]). Figure 3a and b show the measured oscilloscope persistence traces and the photoresponse inter-arrival time histograms of the same devices shown in Figure 2a and b, when biased in avalanche regime (at $I_B = 0.98\ I_{SW}$). Strikingly, we observed that all detectors with a reset time below ~ 1 ns (estimated as in Ref. [2]) showed afterpulsing, as shown in Figure 3a. The afterpulsing also manifested itself in a peak in the inter-arrival time histograms shown in Figure 3b. Although every sub-ns-reset-time device exhibited afterpulsing, the afterpulsing was not observed on every photoresponse pulse [14], as shown by the oscilloscope persistence trace.

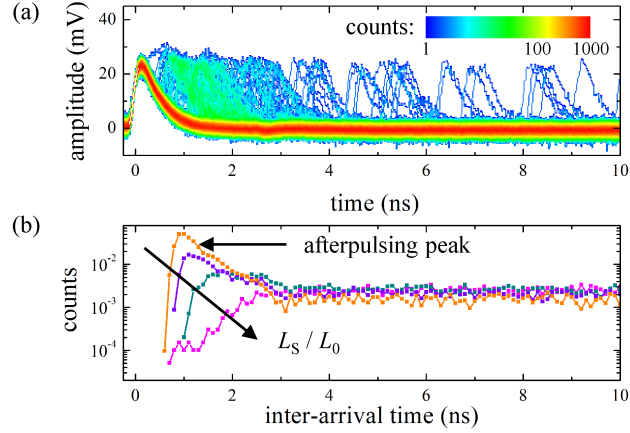

**Figure 3. a.** Oscilloscope persistence map of the response of the 4-SNAP with $r = 1$ of Figure 2a and b. The device was biased in the avalanche regime, at $I_B = 0.98\ I_{SW}$. The color represents the number of occurrences of a certain point (due to overlapping waveforms). Note the logarithmic color scale, and the marked difference in counts between the main (red) region, and after-pulsing (green) regions. **b.** Histograms of the photoresponse pulse inter-arrival time for the same devices used in Figure 2a and b. The devices were biased in the avalanche regime, at $I_B = 0.98\ I_{SW}$.

The afterpulsing we observed on our sub-1-ns-reset-time devices is a different phenomenon from the unstable regime shown in Figure 1 and from the afterpulsing due to the read out circuit [15] (see supplementary material [8]). We attributed the origin of the afterpulsing to the thermal relaxation dynamics of the superconducting nanowires. To support our hypothesis, we used the electrothermal model to simulate the recovery after an HSN event of the initiating section of a 3-SNAPs with the same series inductance as a correctly operating device, as shown in Figure 4a and b, and as an afterpulsing device, as shown in Figure 4c and d.    For both devices, after the normal domain is formed, the nanowire switches back to the superconducting state when its critical current ($I_C$) [16] becomes larger than the current through it ($I_i$). Once the superconductivity is restored, both $I_i$ and $I_C$ relax by increasing towards the steady-state values. While $I_i$ increases with a time constant $(L_0 / 3 + L_S) / R_{load}$, $I_C$ increases at the thermal relaxation rate, which decreases as the nanowire temperature ($T$) approaches the substrate temperature [5, 17].

Therefore, although the relaxation of $I_C$ is initially faster than that of $I_i$, it slows down as $T$ decreases. Figure 4a shows that, if the initial thermal relaxation rate is sufficiently shorter than the electrical relaxation rate, once the superconductivity is restored both $I_i$ and $I_C$ increase towards the steady-state values without crossing again. Therefore, the device fully resets without afterpulsing, as shown in Figure 4b. However, if the thermal and electrical relaxation rates are commensurate and the bias current is close to the steady-state critical current, $I_i$ may exceed $I_C$ again during the recovery, causing the relaxation-oscillation (*RO*) type behavior shown in Figure 4c (i.e. the afterpulsing repeats). The few-nanosecond-long *RO* phase in Figure 4d shows that our simulations could qualitatively reproduce the afterpulsing observed experimentally. We attributed the fact that the device relaxes to the superconducting state after the *RO* phase, instead of latching [18], to the slight imbalance between the current through the initiating and secondary sections (created as a result of the avalanche formation), which prevents all of the sections from latching to the normal state at the same time.

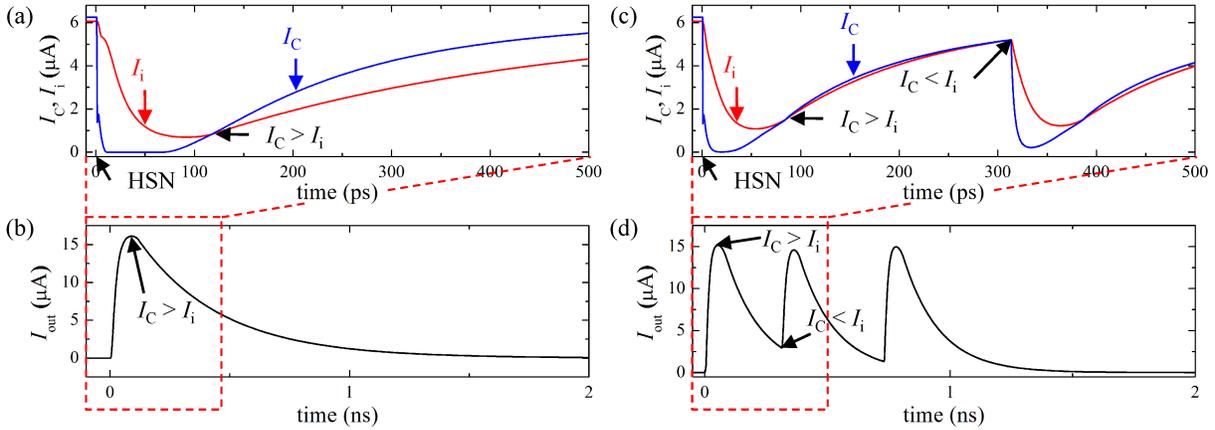

Figure 4. **a.** Simulated recovery after a HSN event (at time 0 s, see black arrow) of the current through the initiating section ($I_i$, in red) of a 3-SNAP and its critical current ($I_C$, in blue). For the simulated device: $L_0 = 13$ nH; $L_S = L_0$ ($r = 0.5$). **b.** Simulation of the output current ($I_{out}$) for the same device of Figure 4a. The red dashed frame encloses the region of the simulation shown in Figure 4a. **c.** Simulated recovery after an HSN event (at time 0 s, see black arrow) of $I_i$ (in red) and $I_C$ (in blue) for a 3-SNAP with $L_0 = 13$ nH; $L_S = 0.25 L_0$ ($r = 2$). **d.** Simulated $I_{out}$ for the same device of Figure 4c. The first pulse starts at time 0 s, when the HSN event occurs in the initiating section. The second and third pulses start at time 310 ps and 730 ps, when $I_i$ exceeds $I_C$ during the current recovery as shown in Figure 4c. The red dashed frame encloses the initial stages of the simulation shown in Figure 4c.

Although we did not observe afterpulsing on any devices with reset times longer than ~ 1 ns, we observed afterpulsing on all of the devices with reset time below ~ 1 ns. We were able to measure afterpulsing-free SNAPs with lower reset times than SNSPDs with the same area (up to a factor of ~ 2 by using 3-SNAPs with $r = 0.5$, see supplementary material [8]), indicating that SNAPs can in principle achieve lower reset time than SNSPDs with the same area. However, the speed of superconducting-nanowire-based detectors appears to be ultimately limited by the nanowire thermal-relaxation dynamics. Indeed, when the electrical- and thermal-relaxation time scales become comparable, superconducting-nanowire-based detectors appear to malfunction by either latching or afterpulsing, depending on the detector architecture and electrical environment [18].

We investigated the speed limit of SNAPs by decreasing the series inductor ($L_S$). As we decreased $L_S$, we observed that: (1) SNAPs with low $L_S$ ($r > 0.1$) emitted trains of current pulses when biased at a lower current than the onset of the arm-trigger regime; (2) the desired decrease in the detector reset time came at the price of an increase in the avalanche current, which decreased the bias range for the correct operation of the devices. Our results indicate that the reset time of SNAPs can be made lower than SNSPDs with the same area by decreasing $L_S$. However, the reset time could not be reduced below ~ 1 ns, as the devices showed afterpulsing. Based on our simulations, we attributed the unstable regime to the rebiasing of the SNAP after an HSN event occurred in one of the sections and the afterpulsing effect to the electrothermal relaxation of the device after an avalanche was triggered. We suspect that the limit on the reset time of ~ 1 ns we observed on our devices was due to the device materials and geometries we employed. Therefore, engineering the thermal environment of the superconducting nanowires (by modifying the substrate material or surface preparation, or by patterning thermally conductive materials on the nanowires) may result in a decrease of the thermal-relaxation time, which is required to allow reducing the detector reset time below 1 ns.

The authors thank: James Daley, Mark Mondol and Prof. Rajeev Ram for technical support. The work by F. Marsili was supported by the Center for Excitonics, under Award Number DE-SC0001088. The work by F. Najafi was supported by IARPA. The work at MIT Lincoln Laboratory was sponsored by the United States Air Force under Air Force Contract #FA8721-05-C-0002. Opinions, interpretations, recommendations and conclusions are those of the authors and are not necessarily endorsed by the United States Government.

# Afterpulsing and Instability in Superconducting Nanowire Avalanche Photodetectors: Supplementary Material

F. Marsili[1], F. Najafi[1], E. Dauler[2], R. Molnar[2], K. K. Berggren[1*]

[1] *Department of Electrical Engineering and Computer Science, Massachusetts Institute of Technology, 77 Massachusetts Avenue, Cambridge, Massachusetts 02139, USA*

[2] *Lincoln Laboratory, Massachusetts Institute of Technology, 244 Wood St., Lexington, Massachusetts 02420, USA*

[*] *corresponding author: berggren@mit.edu*

**Table of Contents**

**Fabrication process and experimental setup**

We fabricated ultranarrow-nanowire (30-nm width, 100-nm pitch) superconducting nanowire avalanche photodetectors with 2, 3 and 4 sections in parallel (2-, 3- and 4-SNAPs) on 5.5-nm-thick NbN films with active area varying from 0.8 to 2.1 µm$^2$ as described in Ref. [1]. We measured photoresponse count rate (*PCR*) and photoresponse inter-arrival time histograms of these detectors in a cryogenic probe station at a temperature of 4.7 K. We used a pulsed gain-switched laser diode (15-ns pulse width; 50-MHz repetition rate; 1550-nm wavelength) for *PCR* measurements and a continuous-wave (CW) laser (1550-nm wavelength) for inter-arrival time measurements. Further details on our experimental setup are reported in Ref. [1].

The photoresponse count rate was estimated as *PCR* = *CR* - *DCR*, where *CR* is the count rate measured when the detector was illuminated and *DCR* is the dark count rate measured when the detector was not illuminated. The photon flux was calculated as $N_{ph} / H$, where $N_{ph}$ is the number of photons per second incident on the device active area and $H$ is a normalization factor [1].

We ensured that the SNAPs operating in avalanche mode were operating in single-photon-detection regime by verifying that the dependence of photoresponse counts on incident power was linear [2].

**Amplitude of the response pulses in the unstable regime**

We measured the histograms of the amplitude of the response pulses of SNAPs biased in the avalanche and unstable regimes. In the avalanche regime, the amplitudes of the detector pulses were distributed around one single average value, which resulted in a histogram with a single peak as shown in Figure SM 1a. In the unstable regime, two sets of pulses were observable (small and large pulses) with two distinct average amplitudes, which resulted in a histogram with two peaks as shown in Figure SM 1b.

Figure SM 1c shows the *PCR* (normalized to the *PCR* at the switching current, $I_{SW}$) vs bias current ($I_B$) curves of a 3-SNAP measured with different trigger levels of the discriminator of the pulse counter. The spurious peak due to the unstable regime significantly decreased with increasing the trigger level (see black arrow). Indeed, as the trigger level was made larger than the average amplitude of the small pulses generated by the device, mostly the large pulses were recorded by the counter. In Figure 1a we intentionally chose to count the pulses of both amplitudes to make the unstable regime evident on the *PCR* vs $I_B$ curves.

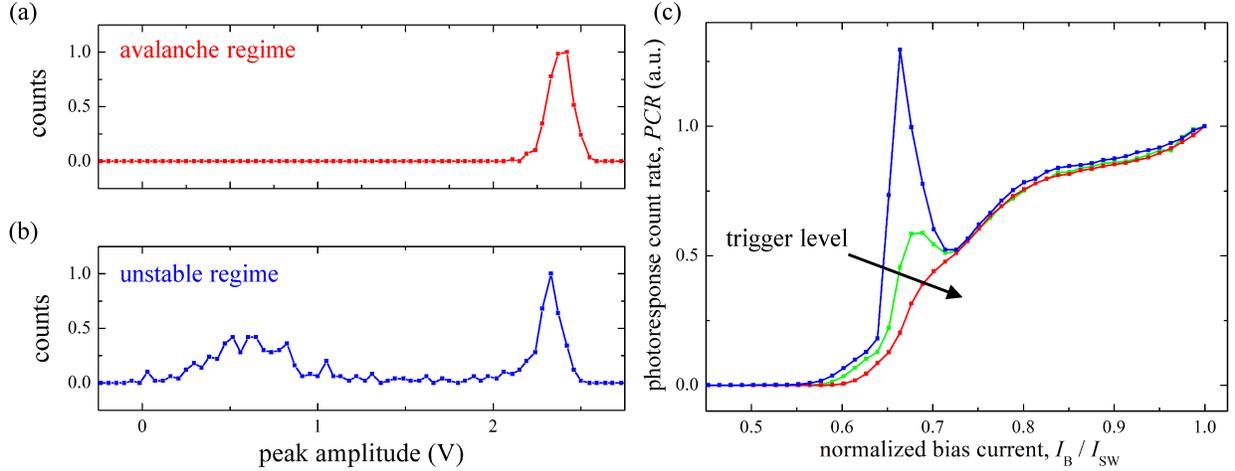

**Figure SM 1. a, b.** Histogram of the photoresponse pulse amplitude of the 3-SNAP of Figure 1a measured in the avalanche ($I_B = 0.9\ I_{SW}$, in red) and unstable ($I_B = 0.65\ I_{SW}$, in blue) regimes. The counts were normalized to the maximum of the histogram. **c.** *PCR* (normalized to the *PCR* at the switching current) vs $I_B / I_{SW}$ of a 3-SNAP with section inductance $L_0 = 13$ nH (estimated from the fall time of the detector response pulse) and series inductance $L_S = 1.8\ L_0$ (redistribution ratio $r = 0.28$). The photon flux on the device active area was $4 \cdot 10^6$ photons per second. The three curves were measured with three different trigger levels: 75 mV (blue), 100 mV (green) and 150 mV (red). The black arrow indicates the direction of increasing trigger level.

## Electrothermal simulation of 2- and 3-SNAPs in the unstable regime

Figure SM 2 shows the electrothermal simulations of the current dynamics of a 2- (Figure SM 2a and b) and a 3-SNAP (Figure SM 2c and d) with $r = 1$, which supports our explanation of the unstable regime.

For the 2-SNAP, after a hotspot nucleation (HSN) event occurs in section 1 (see red curve in Figure SM 2a), no avalanche formed. During the current recovery in the nanowires, the current in section 2 ($I_2$) exceeds the nanowire switching current ($I_2 > I_{SW}$, see blue curve in Figure SM 2a), which starts a periodic sequence of current pulses in the read out, as shown in Figure SM 2b. As we did not include the occurrence of other HSN events in our simulation, the instability cycle repeated for the whole duration of the simulation (~ 40 ns).

For the 3-SNAP, after two subsequent hotspot nucleation (HSN) events occur in sections 1 and 2 (at time = 0 s and 750 ps, Figure SM 2c), no avalanche is formed (unlike what happens in arm-trigger regime [1]) because approximately half of the redistributed current leaks into the read-out resistor $R_{load}$ (as $r = 1$). Once section 2 switches back to the superconducting state, the current through $R_{load}$ ($I_{out}$, Figure SM 2d) starts decreasing and the currents in the SNAP sections increase at the same rate, but from different initial conditions. Indeed, while sections 1 and 2 are depleted of current, section 3 is biased close to $I_{SW}$, as it remained in the superconducting state. When the current in section 3 exceeds the $I_{SW}$ of that section, the different branches of the circuit start swapping their current periodically.

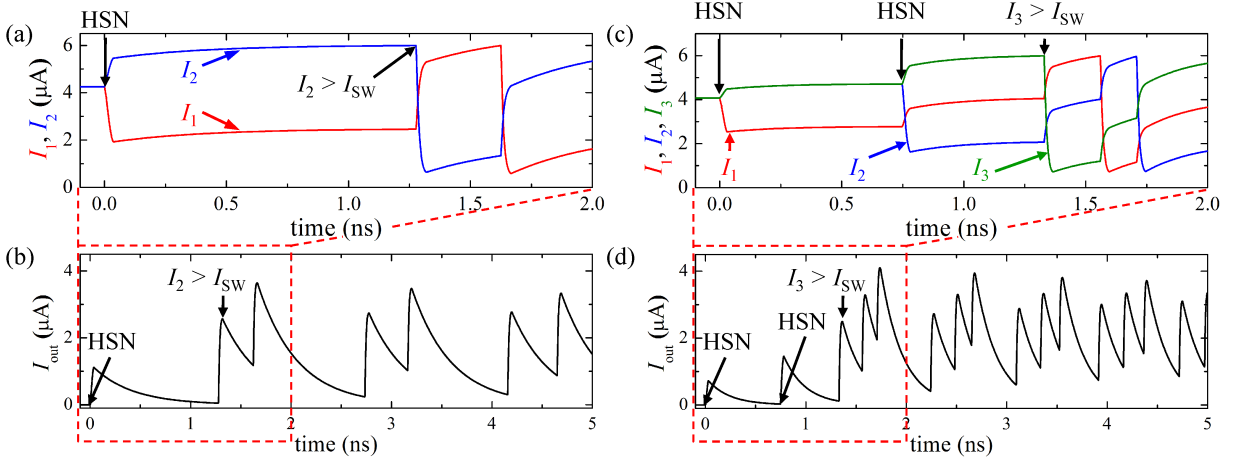

**Figure SM 2. a.** Electrothermal simulation of the currents ($I_1$ in red, $I_2$ in blue) through the sections of a 2-SNAP with $L_0 = 13$ nH and series inductance $L_S = L_0$ ($r = 1$) biased at $I_B = 0.71\ I_{SW}$. Black arrows mark the time at which the HSN event occurs in section 1 (time = 0 s), and the time at which the current through section 2 becomes overcritical ($I_2 > I_{SW}$). **b.** Electrothermal simulation of the current through $R_{load}$ for the same device and conditions used in Figure SM 2a. The red dashed frame encloses the initial stages of the simulation shown in Figure SM 2a. **c.** Electrothermal simulation of the three currents ($I_1$ in red, $I_2$ in blue, and $I_3$ in green) through each of the sections of a 3-SNAP with $L_0 = 13$ nH and series inductance $L_S = 0.5\ L_0$ ($r = 1$) biased at $I_B = 0.68\ I_{SW}$. Black arrows mark the time at which the first HSN event occurs in section 1 (time = 0 s), the second HSN event occurs in section 2 (time = 750 ps) and the current through section 3 becomes overcritical. **d.** Electrothermal simulation of the current through $R_{load}$ for the same device and conditions used in Figure SM 2c. The red dashed frame encloses the initial stages of the simulation shown in Figure SM 2c.

## Dependency of the avalanche current on the number of sections and *r*

In order to have a qualitative estimate of how the avalanche current ($I_{AV}$) depends on the number of sections (*N*) and *r*, we adopted a simplified model of the device operation which assumes that all the current of the initiating section is redistributed to the secondary sections and to $R_{load}$ (so no current remains in the initiating section). The current through the secondary sections after the initiating section switched to the normal state can be written as: $I_s = (I_B / N) + \delta I$, where $\delta I$ is the current redistributed from the initiating section. Assuming that all the current in the initiating section redistributes to the secondary sections and to $R_{load}$, we can write $\delta I = I_B / [N \cdot (N - 1) \cdot (r + 1)]$, which yields: $I_{AV} / I_{SW} = (N - 1) / [N - r / (r + 1)]$ (which assumes that all the sections have the same switching current $I_{SW} / N$). This expression of the avalanche current, obtained by assuming that the current of the initiating section can redistribute to $R_{load}$, is higher than the one calculated assuming perfect redistribution [1] by a factor dependent on $r$.

## Discussion on the afterpulsing of sub-1-ns-reset-time devices

Although when comparing Figure 3a with the inset of Figure 1a the afterpulsing may seem analogous to the unstable regime, the two effects are different for the following reasons: (1) the unstable regime occurred below $I_{AV}$, while the afterpulsing occurred above $I_{AV}$; (2) in the unstable regime the pulses generated by the detectors had two distinct average amplitudes, while the amplitudes of the pulses generated by afterpulsing devices were distributed around one single average value; (3) in the unstable regime the detector emitted pulses continuously, while the afterpulsing had a duration of only ~ 2 ns; and (4) as a consequence of (3), unlike the unstable regime afterpulsing could not be detected in the $PCR$ vs $I_B$ curves because the electrical bandwidth of the pulse counter used for those measurements was only 250 MHz, so the sequence of pulses occurring within less than 4 ns from the first photodetection pulse was integrated and triggered one single count.

The differences between the afterpulsing we measured on our fast-resetting devices and the afterpulsing reported in Ref. [3] on standard superconducting nanowire single photon detectors (SNSPDs) are: (1) the time scale of the afterpulsing in in Ref. [3] is of the order of 100 ns, while the afterpulsing we observed happened within few ns from the main pulse; and (2) unlike what reported in Ref. [3] the afterpulsing we observed did not depend on the detector count rate.

## Low-reset-time SNAPs

By reducing the series inductance ($L_S$), we fabricated SNAPs not affected by afterpulsing with lower reset times than SNSPDs with the same area. Figure SM 3a shows the normalized photoresponse count rate ($PCR$) vs bias current ($I_B$) curve of a 3-SNAP with $r = 0.5$ ($L_S = L_0$). The avalanche current of the device ($I_{AV} = 0.81 I_{SW}$, see black arrow in Figure SM 3a) was only 4 % higher than the avalanche current of 3-SNAPs with $r = 0.1$ ($L_S = 10 L_0$, $I_{AV} = 0.78\ I_{SW}$, see Figure 2c), so the reduction of the series inductance by a factor of 10 did not significantly affect the bias margin of the device. Figure SM 3b and c show the measured oscilloscope persistence traces and the photoresponse inter-arrival time histograms of the same device shown in Figure SM 3a, when biased in avalanche regime (at $I_B = 0.98\ I_{SW}$). The device showed a reset time of 1.9 ns (estimated as in Ref. [1])and no afterpulsing. The total kinetic inductance of the device was $L_{SNAP} = 1.3\ L_0$, which is a factor of 2.3 lower than the kinetic inductance of a SNSPD with the same area ($L_{SNSPD} = 3\ L_0$).

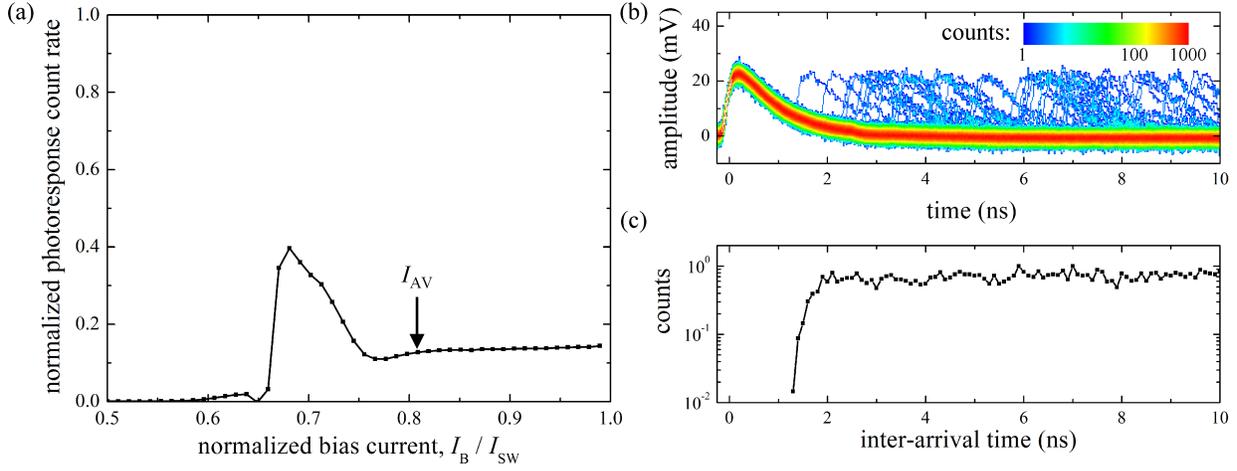

**Figure SM 3. a.** Normalized *PCR* vs $I_B / I_{SW}$ of a 30-nm-wide 3-SNAP (1.47 × 1.47 µm² active area) with $r = 0.5$ ($L_S = L_0$). The *PCR* is normalized to the photon flux (23 Mph/s). The kinetic inductance of one section was $L_0 = 22$ nH (estimated from the fall time of the detector response pulse). The detector avalanche current ($I_{AV}$) was marked by a black arrow. **b.** Oscilloscope persistence map of the response of the 3-SNAP of Figure SM 3a. The device was biased in the avalanche regime, at $I_B = 0.98\ I_{SW}$. The color represents the number of occurrences of a certain point (due to overlapping waveforms). **c.** Histograms of the photoresponse pulse inter-arrival time for the same device used in Figure SM 3a and b. The devices was biased in the avalanche regime, at $I_B = 0.98\ I_{SW}$.